\documentclass[pra,twocolumn,showpacs,amsmath,amssymb]{revtex4}

\newcommand{\beq}{\begin{eqnarray}}
\newcommand{\eeq}{\end{eqnarray}}
\newcommand{\beqq}{\begin{eqnarray*}}
\newcommand{\eeqq}{\end{eqnarray*}}
\newcommand{\w}{\omega}

\newcommand{\g}{\gamma}

\newcommand{\etal}{{\it et al.}}
\newcommand{\ie}{{\it i.e.}}

\def\be{\begin{equation}}
\def\ee{\end{equation}}
\def\bee{\begin{equation*}}
\def\eee{\end{equation*}}
\def\been{\begin{enumerate}}
\def\enen{\end{enumerate}}
\def\beit{\begin{itemize}}
\def\enit{\end{itemize}}

\usepackage{graphicx}
\usepackage{dcolumn}
\usepackage{bm}

\begin{document}
\title{Accurate description of optical precursors and their
relation to \\ weak-field coherent optical transients}

\author{William R. LeFew,$^1$ Stephanos Venakides,$^1$ and
Daniel J. Gauthier$^2$}
\affiliation{Duke University, $^1$Department of Mathematics,
$^2$Department of Physics, Durham, North Carolina 27708 USA}

\date{\today}

\begin{abstract}

We study theoretically the propagation of a step-modulated
optical field as it passes through a dispersive dielectric
made up of a dilute collection of oscillators characterized
by a single narrow-band resonance.  The propagated field is
given in terms of an integral of a Fourier type, which cannot be
evaluated even for simple models of the dispersive dielectric.
The fact that the oscillators have a low number density
(dilute medium) and have a narrow-band resonance allows us to
simplify the integrand.  In this case, the integral can be
evaluated exactly, although it is not possible using this
method to separate out the transient part of the
propagated field known as optical precursors.  We also use
an asymptotic method (saddle-point method) to evaluate the
integral.  The contributions to the integral related to the
saddle-points of the integrand give rise to the optical
precursors.  We obtain analytic expressions for the precursor
fields and the domain over which the asymptotic method
is valid.  When combined to obtain the total transient field,
we find that the agreement between the solutions obtained
by the asymptotic and the exact methods is excellent.
Our results demonstrate that precursors can
persist for many nanoseconds and the chirp in the instantaneous frequency
of the precursors can manifest itself in beats in the transmitted
intensity. Our work strongly suggests that precursors have been observed
in many previous experiments.

\end{abstract}


\maketitle

\section{Introduction}
\label{Sec:Introduction}

A fundamental problem in classical electromagnetism is the propagation
of a disturbance or pulse through a dispersive optical material,
which is characterized by a frequency-dependent complex refractive
index $n(\omega)$. The moment when the field first turns on
(the pulse ``front'') is directly related to the flow of information
propagating through the material \cite{BrillouinBook}.
 The theoretical formulation is straightforward for the case when 
the input field is weak enough
so that the dielectric responds linearly to the applied field.
The calculation of the propagated field $E(z,t)$, assumed to
be an infinite plane wave traveling along the $z$-direction,
involves the evaluation of a Fourier integral that is crucially
dependent on $n(\omega)$ (see Sec.\ \ref{Sec:Theory}). 
The exact evaluation of the integral is impossible even
for simple causal models of the dispersive
optical material.

The first real theoretical headway on the problem was made nearly a century ago by Sommerfeld
and Brillouin (SB), studying a step-modulated input
field, which has zero initial amplitude and jumps instantaneously to a constant
value $A_0$. Many aspects of the solution are similar to those
for other pulse shapes. As summarized in a more recent collection of
their earlier papers \cite{BrillouinBook}, Sommerfeld
and Brillouin were able to show that
the front of the step-modulated pulse always propagates at
the speed of light in vacuum $c$
and hence the flow of information is relativistically causal.
They also found that, after the front and before
the field eventually attains its steady-state value, there exist two
transient wavepackets, now known as the Sommerfeld and
Brillouin precursors. The wavepackets in 
the propagated field arise from contributions to the Fourier integral 
that are localized  near complex frequencies,  known 
as the Sommerfeld or Brillouin ``saddle-points,'' described in
Sec.\ \ref{Sec:Theory}. 
 The concept of precursors exists only over the time in which 
 such localized contributions to the integral occur. 
During this time, the sum 
of the Sommerfeld and Brillouin precursors provides the leading 
asymptotic approximation to the  transient field.
Over the years, various researchers have corrected errors in the SB
calculations as well as extending the work to related problems,
as discussed in great detail by Oughstun and Sherman (OS) \cite{OS}.
It has been suggested that precursors can penetrate
deeper into a material \cite{OS}, which may be of use in
underground communications \cite{choi} or imaging through
biological tissue \cite{bio_precursor}.  

In this paper, we resolve a substantial controversy 
that exists on the observability of
precursors in the optical part of the spectrum. The controversy was 
grounded in the belief  
 that precursors are an ultrafast effect,
where it is difficult to measure directly the
field transients \cite{choi,roberts,alfano,gibson,avenel,
jeong,okawachi}. Under incident frequency and material parameter 
conditions described below, we calculate expressions for 
the precursors and  we find that
their duration  can be long (in the range of nanoseconds) and that 
their duration is controlled by the inverse of the
resonance width.  We also find that the fraction of the transient part of the field that is associated with precursors is controlled by the absorption. 
Furthermore, we reproduce Crisp's formula \cite{crisp} for the total field 
without making an assumption (see discussion and Appendix) that is crucial for Crisp's derivation. 

Our calculations of the precursors and of their
lifetimes assume the following material parameter and frequency conditions. The half-width
at half-maximum of the material resonance $\delta$ is narrow, the carrier
frequency of  the field $\omega_c$ is nearly
equal to the material resonance frequency $\omega_0$, and the
density of oscillators, characterized by the plasma
dispersion frequency $\omega_p$, is small enough so that
certain limits are satisfied, as discussed later. These assumptions result in a
simplification of the Fourier integral for the propagating field
that allows us to carry through the saddle-point calculation explicitly.
Our approximations differ from the approximate theory (and from
the numerical evaluation) of OS toward the calculation of the
saddle-points. The OS theory is
suitable for materials characterized by a broad resonance
and a high density of oscillators, namely the situations
when $\w_p$ and $\delta$ are of the order of $\w_0$.
This restricted range of parameters was originally
considered by SB \cite{BrillouinBook} and used, to a
large extent, by most other researchers investigating
precursor behavior.  The\ OS\ approximations were not
intended for dilute materials with narrow resonance
and lead to unphysical predictions if applied to
this case.  For example, they yield an unphysically large
amplitude of the transmitted field in weak-field
coherent optical transient experiments, recently
performed by Jeong \textit{et al.} \cite{jeong}.

The paper is arranged as follows. In Sec.\ \ref{Sec:Theory}, we
formulate the theory and, in particular, the Fourier integral 
for the propagated field. In Sec.\ \ref{Sec:precursors}, we 
simplify the Fourier integral based on considerations of the
magnitudes of the material parameters and we apply the saddle-point method 
to the simplified integral to obtain highly accurate analytic expressions for 
the saddle-points and for the  Sommerfeld and
Brillouin precursors. We find that the
envelope of both precursors is non-oscillatory and that they
display a frequency chirp.  When these fields are added
to yield the total transient propagated field, we find
that the field envelope oscillates as a result of the chirp. 
In Sec.\ \ref{Sec:analysis}, we derive explicit asymptotic constraints on 
the material parameters that guarantee the accuracy of our approximations used in the simplification of the Fourier integral. We make a direct, exact 
evaluation of this integral in Sec.\ \ref{Sec:Exact}. This calculation can only make predictions concerning the total transient field, but not the individual precursor
fields. It gives identical predictions to the saddle-point theory 
under conditions when the latter is 
valid, as shown in Sec.\ \ref{Sec:Results}. It agrees exactly 
with the result of Crisp obtained with the aid of the slowly 
varying amplitude approximation (SVAA), where we evaluate our
theory in the limit where local field effects are negligible
to make this comparison.
Thus, we demonstrate unambiguously that the
the weak-field coherent optical transients resulting from
the interaction of resonant radiation propagating
through a dilute gas of atoms (\textit{e.g.}, the
$0\pi$ pulse of Crisp \cite{crisp}) consist of optical precursors 
that can persist for many nanoseconds. We conclude that precursors have been
observed in several experiments over the past few decades,
discussed in Sec.\ \ref{Sec:Discussion}. A method for calculating
higher-order corrections to further improve the accuracy of 
our results is given in the Appendix (Sec.\ \ref{Sec:Corrections}).

\section{Theoretical Formulation}
\label{Sec:Theory}
The propagated field $E(z,t)$, assumed to be an infinite plane wave
traveling along the $z$-direction, is expressed as the real part of a
Fourier integral \cite{OS}
\be\label{eq:E}
E(z,t)=-\text{Re} \left[\dfrac{A_0\Theta(\tau)}{2\pi}\int_{-\infty+i0}^{+\infty+i0} \dfrac{e^{\psi(\w)}}{\w-\w_c}d\w
\right],\ee
where the incident field is given by
$E(z=0^+,t)=A_0\Theta(t)\sin(\w_ct)$.  In Eq.\ \eqref{eq:E},
\be \label{eq:rettime}
\tau=t-z/c
\ee
is the retarded time,  $\Theta(\tau)$
is the Heaviside function and 
\be\label{psi0}
\psi(\omega)=i\w\left(\dfrac{zn(\w)}{c}-t\right), 
\ee
where $n(\w)$ is the refractive index at frequency $\w$ and $z$ is the depth of the measurement point.  Equation \eqref{eq:E}
is an exact solution to Maxwell's equations, in integral form, for
a step-modulated field propagating through a dispersive dielectric. 

\begin{figure}[h,t,b]
\includegraphics[scale=.5]{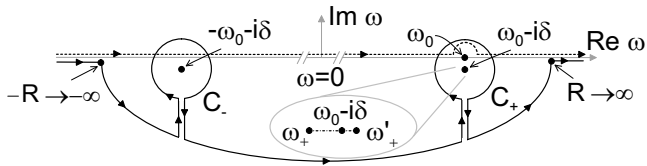}
\caption{\label{fig:omegaplane} The complex $\omega$-plane
showing the original
integration path (dashed line, offset vertically for clarity) and
the deformed contour of integration (solid line, offset vertically
for clarity).  Also shown is the singular point at $\omega=\omega_0-i \delta$
and the simple pole at $\omega=\omega_0$.}
\end{figure}

In the original work of SB \cite{BrillouinBook}, and for much
of the later work including that of OS \cite{OS}, the dielectric is
modeled as a collection of damped harmonic oscillators
that fills the half space $z \ge 0$ (known as a Lorentz dielectric).
This model assumes that the density of oscillators (proportional
to $\omega_p$) is not too large.  For high densities,
the local field about an oscillator has a substantial contribution
from its neighboring oscillators. Such local-field effects can
be taken into account using the Lorentz-Lorenz formulation of
the complex refractive index, which is given by \cite{Oughstun2003,Oughstun2003a}
\begin{equation}\label{refindex0} n(\omega)=\left(1-\dfrac{\omega_p^2}{a}\right)^{1/2}
\end{equation}
where
\begin{equation} \label{eq:refindexden}
a=\omega^2-\omega_0^2+2i\omega\delta+\frac{1}{3}\w_p^2. \end{equation}
The standard expression for the refractive index that does not
take into account local-field effects can be obtained by dropping
the last term in Eq.\ (\ref{eq:refindexden}).
The refractive index depends on the frequency analytically except at 
the  complex frequencies at which the square root has a branchpoint.  
This occurs when the quantity under the radical is either infinite, 
in which case $a=0$, with roots  
\be \label{Bbranch}\w_\pm= -i\delta\pm\sqrt{\w_0^2-\frac{1}{3}\w_p^2-\delta^2},\ee
or when it is zero, with $a=\w_p^2$, and roots 
\be \label{Sbranch}\w_\pm'= -i\delta\pm\sqrt{\w_0^2+\frac{2}{3}\w_p^2-\delta^2}. \ee
For the material parameters of interest, the four roots lie in the lower complex half-plane, 
symmetrically about the imaginary axis, with imaginary part $-i\delta$. 
They are connected by two branchcuts, as shown in Fig. \ref{fig:omegaplane}.

In order to facilitate the evaluation of the complex integral for the propagated field, we deform the original contour of integration (\ie, the real axis, slightly moved to leave the pole under it), to a semicircle of infinite radius in the lower complex half-plane that connects the $\pm$ real points  at infinity. The value of the integral over this semicircle is zero. As the deforming contour cuts through the two obstructing branchcuts,   it leaves behind two clockwise-oriented contour loops, $C_+$ encircling the right branchcut and $C_-$ encircling  the  left one. As  shown in Fig. \ref{fig:omegaplane}, 
the loops, except for their orientation,  are chosen to be  mirror images of each other with respect to the imaginary axis.  Furthermore, $C_+$ is chosen to pass through the Sommerfeld and Brillouin 
saddle-points ({\it i.e.}, stationary points of the exponential
in the Fourier integral, see below)
in the right half-plane.  This implies that $C_-$ also passes through the two 
saddle-points in the left half-plane. Within the regime of validity of the saddle-point method, the main contributions to the integrals are localized near these 
saddle-points.  The pole residue contribution must be
added to the field if the pole $\w_c$ is positioned outside $C_+$. 

The integral over $C_-$, representing the so-called counter-rotating contribution to the propagated field, 
can be efficiently represented as an integral over $C_+$ through a   
change of the variable of integration $\w\to-\overline\w$ (bar indicates 
complex conjugate) and using the symmetry of the exponent,
\be \psi(-\overline\w)=\overline{\psi(\w)}.\ee
We thus obtain
\be \label{eq:crotate}
\int_{C_-}\dfrac{e^{\psi(\w)}}{\w-\w_c}d\w
=-\overline{\int_{C_+}\dfrac{e^{\psi(\w)}}{\w+\w_c}d\w}.
\ee 
Inserting Eq.\ (\ref{eq:crotate}) in Eq.\ \eqref{eq:E}, we obtain  
\be \label{eq:E1}
E(z,t)=E_{rot}(z,t)+E_{crot}(z,t)+\xi_{C_+}E_c,
\ee
where the rotating term is given by 
\be\label{Erot}
E_{rot}(z,t)=-\text{Re} \left(\dfrac{A_0\Theta(\tau)}{2\pi}
\int_{C_+}\dfrac{e^{\psi(\w)}}{\w-\w_c}d\w
\right),\ee
the counter-rotating term is 
\be\label{Ecrot}
E_{crot}(z,t)=\text{Re} \left(\dfrac{A_0\Theta(\tau)}{2\pi}
\overline{\int_{C_+}\dfrac{e^{\psi(\w)}}{\w+\w_c}d\w}
\right),\ee
the pole contribution is
\be
\label{svamain} 
E_c=A_0\Theta(\tau)\mbox{e}^{-\alpha_0 z/2}\sin \left(\w_0\tau+\frac{\w_0 \Delta n_r}{c}z \right),
\ee
which represents the main field,  
and $\xi_{C_+}=0 \ \mbox{or} \ 1$ according to whether $C_+$ encloses the pole 
or not (if the pole is enclosed, the main field is 
contained in $E_{rot}$).  The sum of the rotating term and the third term in Eq.\ (\ref{eq:E1}) remains constant under 
deformations of the contour $C_+$.

The expression for the main field (\ref{svamain}) displays exponential attenuation as a function of propagation distance, which is governed by the absorption coefficient $\alpha_0$ at frequency $\w_c=\w_0$.  It is defined through the relation 
\be \label{eq:absorption}
\alpha_0 = \frac{2 \w_0 n_i(\w_0)}{c},
\ee
where 
\be 
n_i(\w_0)=\text{Im}[n(\w_0)].
\ee
The value of the refractive index at the resonant frequency is given by, 
\be n(\w_0)=\sqrt{\frac{1+ i m/3}{1-i m/6}},\ee
or, in polar form, 
\be n(\w_0)
=2\sqrt{\frac{m^2+9}{m^2+36}}\text{e}^{i [\tan^{-1}9m/(18-m^2)]/2},\ee  
where 
\be m=\frac{\w_p^2}{\w_0\delta}. \ee
Thus, 
\begin{eqnarray} \label{ni}  n_i(\w_0)&=&\text{Im}[n(\w_0)] \\
&=& 2\sqrt{\frac{m^2+9}{m^2+36}}
\sin\left(\frac{1}{2}\tan^{-1}\frac{9m}{18-m^2} \right)\nonumber \end{eqnarray}
and 
\begin{eqnarray} \label{nr}n_r(\w_0)&=&\text{Re}[n(\w_0)] \\
&=& 2\sqrt{\frac{m^2+9}{m^2+36}}
\cos\left(\frac{1}{2}\tan^{-1}\frac{9m}{18-m^2} \right).\nonumber \end{eqnarray}
The main signal also experiences a $z$-dependent phase shift arising from the real part of the refractive index, where $\Delta n_r = n_r(\w_0)-1$.  As discussed in Sec.\ \ref{Sec:Discussion}, our derivation of expressions for optical precursors allows for large absorption coefficients and Eqs. (\ref{ni}) and \eqref{nr} must be used.  On the other hand, a simplified expression for the refractive index can be obtained, and a connection to other treatments of optical pulse propagation can be made when the absorption is small ($\alpha_0 \ll \w_0/c$).  In this case,
\be
n_i(\w_0) \simeq \frac{\w_p^2}{4\w_0 \delta},
\ee
\be
n_r(\w_0) \simeq 1,
\ee
\be
\alpha_0 \simeq \frac{\w_p^2}{2 c \delta},
\ee
\be
\Delta n_r \simeq 0.
\ee
Here, we see immediately that $\alpha_0$ scales with $\w_p^2$ and inversely with $\delta$.

Using Eqs.\ (\ref{Erot})-(\ref{svamain}) in the expression for the field (\ref{eq:E1}), we calculate the precursors from contributions from only 
the Sommerfeld and Brillouin saddle-points in the right 
half-plane; they include the contributions from their symmetric counterparts 
in the left  half-plane through the second integral. Henceforth, references 
to saddle points or a branchcut are to the ones in the right half-plane.

In order to perform the calculation of the saddle-points explicitly, and 
thus obtain an explicit expression of the precursor fields, we 
focus on an asymptotically large material resonance frequency $\w_0$. 
In this limit, the material is dilute ($\w_p\ll\w_0$) and narrowbanded 
($\delta\ll \w_0$). In the scale of $\w_0$,  the branchpoints  
\eqref{Bbranch}  and \eqref{Sbranch} of the right half-plane ($\w_+$ and $\w_+'$, respectively)  collapse asymptotically to the ``singular point'' $\w_0-i\delta$, as illustrated
in Fig.\ \ref{fig:omegaplane}. The more precise asymptotic formula for the midpoint 
of the collapsing branchcut (not needed in our calculation) is $\w_0-i\delta+(\w_p^2/12-\delta^2/2)/\w_0$. The length $l$ of the branchcut, \ie, the difference $\w_+'-\w_+$, is given asymptotically by
\be\label{l}
l\sim\frac{\w_p^2}{2\w_0}\ll \w_0.
\ee

We make the singular point the center of a new frequency variable, denoted by $\w_*$ and defined by 
\be\label{w*}
\w=\w_0-i\delta+\w_*.
\ee 
We seek parameters (frequency, material, depth, retarded time) for which the values of   $\w_*$ 
at the Sommerfeld and Brillouin saddle-points are at a scale that is intermediate 
between the large material frequency $\w_0$ and the 
small length $l$ of the branchcut. We require 
\be \label{scaling}\frac{\w_p^2}{2\w_0}\ll \w_*\ll \w_0. \ee 
As a result, the saddle-points view the branchcut as a point.
Furthermore, for near resonance excitation ($\w_c \simeq \w_0$),
the saddle points are disproportionately farther away from the center
frequency $-\w_0+i\delta$ of the \textit{counter-rotating} term
than from the center frequency $\w_0+i\delta$ of the rotating terms.
Using these facts allows us to obtain explicit expressions for the
precursors and for the transient field, as discussed below.

\section{Calculation of the precursors}
\label{Sec:precursors}
In order to calculate the integrals that give the field, we seek to isolate  
the dominant terms in the exponent $\psi$ and, in particular, in the expression 
for the refractive index. 
The value of $a$ in Eq. \eqref{refindex0},  
expressed in terms of the new frequency variable $\w_*$, is 
\be\label{a} a=2\w_0\w_*+\w_*^2+\frac{1}{3}\w_p^2.\ee
Following our scaling, 
\be  a\sim 2\w_0\w_*,\ee   
and the term $\w_p^2/a$ in Eq. \eqref{refindex0} satisfies 
\be \frac{\w_p^2}{a}\sim\frac{\w_p^2}{2\w_0\w_*}\ll 1.\ee 
Defining 
\be\label{refindex2} n_1=\frac{\w_p^2}{4\w_0\w_*},\ee
we write the refractive index as 
\be\label{refindex1} n=1-n_1+d, \ee 
where the error term $d$ satisfies $d=O(n_1^2)$.
We insert Eq.\ (\ref{refindex1}) in Eq.\ \eqref{psi0} for $\psi$ to obtain
\be \psi=-i\w\tau-i\frac{z}{c}\w(n_1-d). \ee
where we have used the retarded time (Eq.\ (\ref{eq:rettime})).

\begin{figure}[h,t,b]
\includegraphics[scale=1]{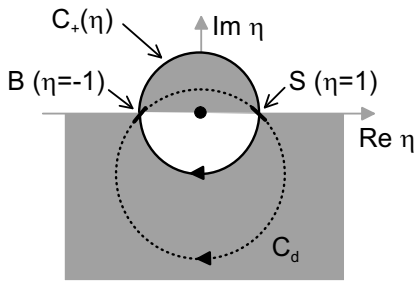}
\caption{\label{fig:eta0} The complex $\eta$-plane showing
contour of integration $C_d$ (dashed line) and the short stretches
along $C_d$ that give the dominant contribution to the integral. The
contour $C_d$ is a deformation of the contour $C_3$ (not shown
for clarity).  The
solid line is a circle of radius 1 centered at the origin.  The shaded
regions are the portion of the $\eta$-plane for which the real part
of the exponent of the integrand is negative.}
\end{figure}

We insert the change of variable \eqref{w*} into this 
expression and arrange the terms into three groups, 
as indicated here with the aid of braces 
\begin{eqnarray} \label{psi} 
\psi=\left\{-i\w_0\tau-\delta\tau\right\}
+\left\{-i\w_*\tau-i\frac{z}{c}\w_0n_1\right\} \\
+\left\{i\frac{z}{c}\w_0d-\frac{z}{c}\delta(n_1-d)
-i\frac{z}{c}\w_*(n_1-d)\right\}.   \nonumber 
\end{eqnarray}
The two terms of the first group are independent of the variable of integration and give the leading amplitude and phase contributions to the integral. Our theory applies to parameters 
(to be identified below) for which the second group is dominant, allowing  
for the third group, labeled $\psi_{rem}$ (mnemonic remainder) to be neglected in the calculation of the saddle-points. The second group is further simplified by introducing the rescaled frequency  
variable $\eta$, defined by 
\be\label{eta} \w_*=\frac{q}{\tau}\eta,\ee
where $q$ is  given by 
\be \label{que} q=\frac{\w_p}{2}\sqrt{\frac{z\tau}{c}}.\ee 
Using these notations, the phase is given by    
\be \label{psieta}  \psi=\{-i\w_0\tau-\delta\tau\}+\{-iq(\eta+\eta^{-1})\}+\psi_{rem}.  \ee
Inserting this expression into Eqs. 
\eqref{Erot} and \eqref{Ecrot} and changing the variable of integration to $\eta$, we obtain  
\be\label{Erot2}E_{rot}(z,t)=-\text{Re} \left[\frac{A_1}{2\pi}\oint_{C_+(\eta)} \mbox{e}^{\psi_{rem}}\dfrac{\mbox{e}^{-iq(\eta+\eta^{-1})}}{\eta-i\sigma}d\eta\right],
\ee
\be\label{Ecrot2}
E_{crot}(z,t)=\text{Re} \left[\frac{A_1}{2\pi
}\oint_{C_+(\eta)}\mbox{e}^{\psi_{rem}} \dfrac{\mbox{e}^{-iq(\eta+\eta^{-1})}}{\eta-i\sigma+2(\w_c\tau/q)}d\eta\right],\ee
\be\label{A1}A_1=A_0\Theta(\tau)\mbox{e}^{-\delta\tau-i\tau\w_0},
\ee
\be \label{sigma1}\sigma=2\frac{\delta}{\w_p}\sqrt\frac{c\tau}{z},\ee
where ${C_+(\eta)}$ is the image of the contour $C_+$ in  the $\eta$ plane as shown in
Fig.\ \ref{fig:eta0}. 
We note that formulae \eqref{Erot2} and \eqref{Ecrot2} are {\it exact}, in spite of their 
derivation being guided by asymptotic considerations. 

We now turn to our calculation of the precursors, which implements the following approximations.  
\begin{enumerate}
\item We neglect the terms of  $\psi_{rem}$ altogether, \ie,\ we set 
$\psi_{rem}=0$ (see discussion in section \ref{Sec:analysis}).  The saddle-points are then approximated as the stationary points of $\eta+\eta^{-1}$, namely, 
$\eta=1$ for the Sommerfeld precursor and $\eta=-1$ for the Brillouin precursor (see Fig.\ \ref{fig:eta0}). 
The saddle-point approximation of the integrals is 
straightforward and given below. The omitted terms 
give only higher order contributions to the phase and 
amplitude of each precursor. A method for obtaining higher-order
corrections is discussed in the Appendix.
\item The counter-rotating term of the field $E_{crot}$ is neglected.
In our case of near-resonant excitation and with our scaling, the relative error  introduced is of order  $q/(\tau\w_0)=(\w_p/\w_0)\sqrt{z/(c\tau)}\ll 1$  (see relation \eqref{n1-subd} below), following the fact that  $|\w_*|=q/\tau$ at the saddle-points. 
\end{enumerate}
As a result of these approximations, the propagated field is given as 
\be \label{Eapprox}
E(z,t)=-\text{Re} \left[\frac{A_1}{2\pi}\oint_{C_+(\eta)} \dfrac{\mbox{e}^{-iq(\eta+\eta^{-1})}}{\eta-i\sigma}d\eta\right]+\Theta(\sigma-1),
\ee
where $C_+(\eta)$ is the unit circle in the $\eta$ plane (see Fig.\ \ref{fig:eta0}). 

The saddle-point method for evaluating the integral 
in Eq. \eqref{Eapprox} requires {\it large values} of  $q$, the relative error of 
its approximation to the value of the integral being of order $1/q$.
In order to apply the method, we deform the contour of integration $C_+(\eta)$  to the contour $C_d$ (see Fig. \ref{fig:eta0}), oriented clockwise, that passes through the saddle-points $\eta=\pm 1$ cutting the real axis at angle $\pi/4$.
The value that the real part of the exponent   assumes at 
the saddle-points is maximal, in comparison to its values in the 
shaded region shown in the figure.  The contour of integration $C_d$
 passes through the saddle-points and stays in the shaded region.
In the limit of large $q$, 
exponential decay in the shaded region makes the contribution of 
the part of the contour close to the saddle points dominant. 
The main contributions to the integral arise from the two short stretches of
the contour $C_d$ in the neighborhood of the saddle points, 
which have been  chosen in
the steepest descent direction ($\pi/4$ angles with the real axis) where the exponential is purely real and the integrands
are approximated by Gaussians. Large values of $q$ localize the Gaussians
at the saddle-points. Thus, the length of the stretches tends to zero as $q$ increases. The
contributions from the two saddle-points to the rotating term of the field, 
obtained from the exact calculation of the Gaussian integrals, are
 \beq\label{somprec}
E_S(z,t)&=&\mbox{Re}\left[\dfrac{iA_1\mbox{e}^{-i(2q-\frac{\pi }{4})}}{2\left({\pi q}\right)^\frac{1}{2}(1-i\sigma)}\right],\\ \label{brilprec}
E_B(z,t)&=&\mbox{Re}\left[\dfrac{iA_1\mbox{e}^{i(
2q-\frac{\pi }{4})} }{2\left({\pi q}\right)^\frac{1}{2}(1+i\sigma)}\right],
\eeq
with the subscript $S \ (B)$ for the Sommerfeld (Brillouin) precursor field. Inserting Eq.\ \eqref{A1} into these obtains the
two precursor fields
\beq\label{somprec1}
E_S(z,t)&=&\mbox{Re}\left[\dfrac{iA_0\Theta(\tau)
\mbox{e}^{-\delta\tau-i(2q-\frac{\pi }{4})}}{2\left({\pi q}\right)^\frac{1}{2}(1-i\sigma)}
\mbox{e}^{-i\tau\w_0}\right],
\\ \label{brilprec1}
E_B(z,t)&=&\mbox{Re}\left[\dfrac{iA_0\Theta(\tau)
\mbox{e}^{-\delta\tau+i(2q-\frac{\pi }{4})} }{2\left({\pi q}\right)^\frac{1}{2}(1+i\sigma)}
\mbox{e}^{-i\tau\w_0}\right].
\eeq

\begin{figure}[t,b]
\includegraphics[scale=1]{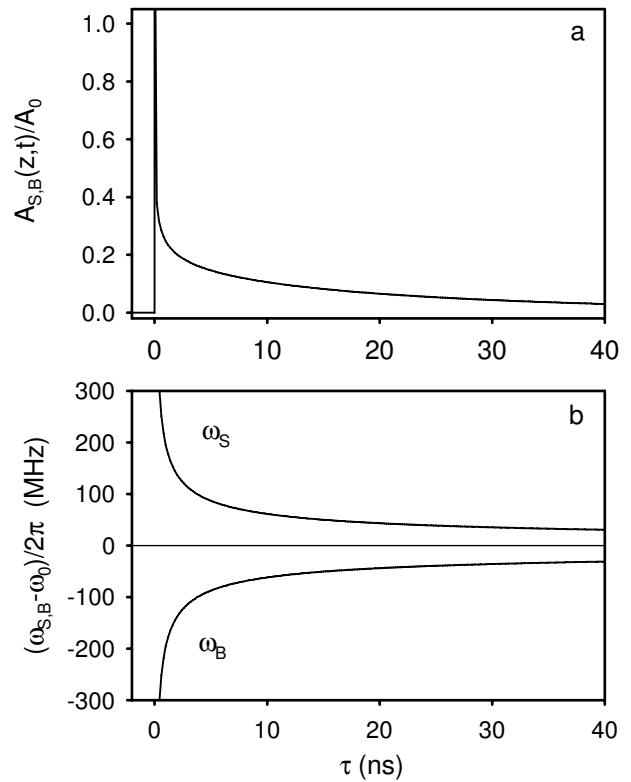}
\caption{\label{fig:precursors} a) The envelope and b) frequency of the Sommerfeld and Brillouin precursors field with the medium parameters of Ref. \cite{jeong} and $z=20$ cm.}
\end{figure}

The precursors display rapid oscillations at a frequency
close to $\omega_0$,  modulated by a complex-valued envelope. Both
precursor envelopes have the same modulus
\be\label{precursampl}
A_{S,B}(z,t)=\dfrac{A_0\Theta(\tau)\mbox{e}^{-\delta\tau}}{2\left({\pi q(1+\sigma^2)}\right)^\frac{1}{2}}.
\ee
The precursors decay exponentially with time constant $1/\delta$, supporting our statement that they persist
for a time determined by the resonance half-width, which can be in the nanosecond time scale for a dilute gas of cold atoms
\cite{jeong}, for example. The precise value of the frequency of the Sommerfeld and Brillouin precursors is determined by
taking the derivative of the phase $\tau\w_0+2q$ with respect 
to $\tau$ in Eqs. \eqref{somprec1} and \eqref{brilprec1},
respectively, yielding
\be\w_S=\omega_0+\frac{q}{\tau}, \ \ \ \w_B=\omega_0-\frac{q}{\tau}.\ee
Note that the precursors frequencies are equal to the real part
of the respective saddle points in the complex $\omega$-plane.
Figure \ref{fig:precursors} shows the envelope and frequencies of the
precursors using the materials parameters of the experiment of
Jeong \etal \ \cite{jeong} with
 $\w_0=2.5\times10^{15}$ $s^{-1}$, $\w_p=3\times10^9$ $s^{-1}$, $\delta=3\times10^7$ $s^{-1}$, $\w_c=\w_0$, but with a longer
medium length $z$=20 cm.  It is seen that the envelop persists
for many nanoseconds and that the precursor frequencies are
within a few hundred MHz of the resonance within a
few nanoseconds.

The total transient field is approximated by the sum of the two precursor fields
\be E_T(z,t)=E_S(z,t)+E_B(z,t). \ee
Because the individual precursors are chirped, $E_T$ will display oscillations
at the beat frequency $\omega_S-\omega_B=2q/\tau$, whose period increases with $\tau$ (the beat frequency decreases with $\tau$).  The
envelope decays with time constant $ 1/\delta$ as do the individual precursors.
The expression for the envelope of the total precursor field is given by
\be\label{totalenvelope}
A_T(z,t)=2A_{S,B}(z,t)\textnormal{cos}(2q-\frac{\pi}{4}-\textnormal{tan}^{-1}\sigma),
\ee
where $q$ and $\sigma$ are given by \eqref{que} and \eqref{sigma1}, respectively.
The pole, located on the imaginary axis at $i\sigma$, starts at the origin when $\tau=0$
and moves up the imaginary axis as time increases. 
Its contribution, \ie, the main field $E_c$, is negligible 
compared to the precursor field when  
the pole is in the shaded region. Outside the shaded region, 
the main field is dominant.

%

\section{Parameter Analysis}
\label{Sec:analysis}

The determination of the range of parameters for which our calculation of the precursors
is accurate follows from assumptions we have made, which we now summarize.

\begin{enumerate}
\item The {\it requirement of small} $n_1$
in the expression for the refractive index \eqref{refindex1} gives 
\be \frac{\w_p^2}{2\w_0\w_*}\ll 1,  \ee
or 
\be \frac{\w_p^2}{2\w_0}\ll \w_*.   \ee
This constraint is identical to the left relation \eqref{scaling}.  
Inserting $\w_*=q/\tau$, we obtain, after simple algebra, 
\be \frac{\w_p}{\w_0}\ll \sqrt{\frac{z}{c\tau}}.\ee
\item The {\it requirement of large} $q$ yields the constraint 
\be \w_p \gg 2\sqrt{\frac{c}{z\tau}}, \ee
which can be obtained from the definition of $q$.
\item The {requirement of the asymptotic vanishing} of the  
second term of the $\psi_{rem}$ (third group  of terms of Eq. \eqref{psi}),
\be \frac{z}{c}\delta(n_1-d)\ll 1,\ee
which, thus, does not contribute to the leading order of the integral, 
partly justifying our neglecting of $\psi_{rem}$.
The requirement, expressed  in terms of material parameters, is
\be\label{small2} \frac{\w_p\delta}{\w_0}\sqrt{\frac{z\tau}{c}}\ll 1.\ee
\item The {\it dominance requirement} constrains the remaining 
(first and third) terms of $\psi_{rem}$  to be significantly smaller than the terms of the second group along the contour of integration. 
Since the two terms of the dominant group have comparable magnitudes, it suffices to make the comparison with only the second term in the dominant group. The ratios of the first and third terms of $\psi_{rem}$ 
by the second term of the dominant group are, respectively, 
\begin{align}
\frac{d}{n_1}
\sim\frac{\w_p^2}{2\w_0\w_*}
\sim \frac{\w_p}{\w_0}\sqrt{\frac{c\tau}{z}}\ll 1, \\
\frac{\w_*}{\w_0} 
\sim \frac{\w_p}{2\w_0}\sqrt{\frac{z}{c\tau}}\ll 1
.\end{align}
These  relations  are identical to the scaling (left and right, respectively) of relation \eqref{scaling}, which is thus satisfied automatically as a result of the  requirement.  

By neglecting these two terms  of $\psi_{rem}$, while falling short of requiring their asymptotic vanishing,  we lose a {\it phase} term in the leading order expression for the precursors. That the error made is only in the phase follows from the fact that, after setting the second term of $\psi_{rem}$ equal to zero, the saddle-points are real (in the $\eta$ variable) and the exponent is purely imaginary. The dominance requirement guarantees that the error in the phase is of higher order compared to the phase correction from the second group in $\psi$ that produces the chirp. In terms of physical insight 
gained by our result, tolerating this error is preferable to further constraining 
the material parameters. The dominance requirement also guarantees
there is no other leading order error in the application of the saddle-point method.
\end{enumerate}

Collecting the independent constraints leaves us with 
\be
\label{largeq}2\sqrt{\frac{c}{z\tau}}\ll \w_p\ee
for large $q$ (the requirement of a large value of $q$ is modest - the saddle-point method
already gives a quite good approximation to the value of the integral for a value of $q$ of $3$ or $4$);
\be\label{n1-subd}\frac{\w_p}{\w_0}\ll \sqrt{\frac{z}{c\tau}}\ee
for $n_1\ll 1$ and dominance of the second group in $\psi$; and
\be\label{subd1}\frac{\w_p}{\w_0}\ll  \sqrt{\frac{c\tau}{z}},\ee
and 
\be\label{small2} \frac{\w_p\delta}{\w_0}\sqrt{\frac{z\tau}{c}}\ll 1.\ee
for the asymptotic vanishing of the second term of $\psi_{rem}$.

When these constraints are satisfied, (a) the precursors exist at the specified  
retarded time (\ie, the calculation 
of the transient field as the sum of saddle-point contributions applies) and (b) our calculation of the precursors is accurate. 
When some constraint is violated, one of these statements may not be true. 
 Assuming suitable fixed frequency, 
depth and material parameters, constraints \eqref{largeq} and \eqref{subd1} can be satisfied 
only past a (usually short) retarded time $\tau$. The constraints are satisfied 
in a time-range  beyond this, until, for time large enough, constraint \eqref{n1-subd} is necessarily violated and our method loses accuracy.

The upper time-limit of validity of our constraints may be overshadowed by the 
additional practical constraint that $\delta\tau$ must be fairly small 
for the precursor to be observable. When 
\label{observability}\be \frac{\delta^2}{\w_p^2}=\frac{z}{c\tau},\ee
the real exponentials multiplying the amplitudes of the precursor
and of the main field, respectively, are equal. To be solidly 
in the regime where the precursor dominate over the main field, we require
 \be \frac{\delta^2}{\w_p^2}\ll\frac{z}{c\tau}.\ee

Generally, when relation \eqref{largeq} is comfortably satisfied, 
but some other constraint fails, we expect that the precursors exist, but our calculation 
loses accuracy. To gain accuracy, we apply a corrective scheme described in the Appendix.  

\section{Exact Evaluation of the Integral for the Approximate Field}
\label{Sec:Exact}
Equation \eqref{Eapprox} gives the approximate propagated field, 
in which $\psi_{rem}$ and the counter-rotating terms have been ignored (the same approximations used to obtain the expression for the precursor fields). In order to make an exact calculation of the integral in this equation, we change to an angle variable of integration defined through $\eta=\exp[i(\rho+\pi/2)]$.
We thus obtain
\be\label{svafield11}
E(z,t)=\text{Re} \left[\frac{iA_1}{2\pi
}\int_{-\pi}^{\pi} \dfrac{\mbox{e}^{2iq\sin\rho}}{1-\sigma\mbox{e}^{-i\rho}}d\rho\right]
+\Theta(\sigma-1)E_c.
\ee

When the integrand of \eqref{svafield11} is expanded in a series of
powers of $\sigma<1$, each of the integrals in the series represents a Bessel function and the total
field is given by
\be
E(z,t)=
\textnormal{Re}\left[iA_1 \sum_{k=0}^\infty \sigma^kJ_k(2q)\right],\label{exact}
\ee
where $J_k$ is the Bessel function of order $k$. The 
approximate transient field is
obtained when the pole-contribution to the field (\ref{svamain}), representing the main field, is subtracted from Eq.\ (\ref{exact}). While the ensuing expression for the transient field   is exact in the limit considered here, it does not allow separating the two precursor fields and does not even make a statement
about the existence of individual precursors. This is due 
to the fact that  the definition of the  precursors is tied to the 
application of the saddle-point method in the calculation of the field. 

A similar calculation is possible when $\sigma>1$. In this case, the
expansion is in powers of $\sigma^{-1}$ and the final formula is
again a series of Bessel functions. Although we only have treated
the case of a resonant field ($\omega_c=\omega_0$), the result for a near resonant field is
obtained by inserting an imaginary part with the
frequency difference in $\sigma$ and expanding in powers of $\sigma$ or $\sigma^{-1}$ according to whether $|\sigma|$ is less than or greater than unity.

In the parameter range of negligible local field effects ($\alpha_0 \ll \w_0/c$), 
an alternative formula for $\sigma$ is 
\be \sigma=\sqrt{\frac{\delta \tau}{\alpha_0z/2}}. \ee
Interestingly, when this expression for $\sigma$ is used, Eq.\ (\ref{exact}) is \textit{identical} to that found by
Crisp who used the slowly varying amplitude approximation (SVAA) (see Eq.\ (29) of Ref.\ \cite{crisp}). We derive Eq.\ (\ref{exact})
only with assumptions about the material properties; we
make no assumption about the slowness of the variation of
the electromagnetic field. Our formula (\ref{exact}) is still valid, even when the 
relation $\alpha_0 \ll \w_0/c$ is violated, as long as $\sigma$ is 
defined by Eq. \eqref{sigma1}.

\section{Results}
\label{Sec:Results}
For the first time, we have derived analytic expressions for
optical precursors for a material with a narrow resonance
and a low oscillator number density (small plasma frequency). 
Our formulae are valid within explicitly specified sub-ranges 
of the space-time range of the existence of precursors. The latter consists 
of the range of points in space-time over which, field contributions to the main Fourier integral (Eq.\ (\ref{eq:E})) are localized at isolated saddle-points of the complex frequency plane. Our precursor theory limits itself to such saddle 
points that are sufficiently close to the resonant frequency for the 
counter-rotating field contributions to be negligible and sufficiently far from it 
to allow approximating the branchcut in the refractive index by a singular point.  
These restrictions apply to the long tail of the precursors, \ie\ after the narrow front of the transient wave has passed. The saddle-points involved are isolated. The front (it displays degenerate or near-degenerate saddle-points) and the very early tail, both requiring the counter-rotating contributions, are not addressed in this study. The exact expression of the field is written in the form of Eqs. \eqref{Erot2}, 
\eqref{Ecrot2}, \eqref{A1}, \eqref{sigma1}. The approximate field is given by Eq. \eqref{Eapprox}, on which the saddle-point method is performed to yield 
the precursors \eqref{somprec1} and \eqref{brilprec1}. The space-time and 
material parameter constraints that define the 
range of validity of the derivation of the approximate field  are given by the relations \eqref{largeq}, \eqref{n1-subd}, \eqref{subd1} and \eqref{small2}. 
The exact evaluation of the approximate field \eqref{Eapprox}, in terms of Bessel 
functions is given by Eq. \eqref{exact}. 

\begin{figure}[t,b]
\includegraphics[scale=1]{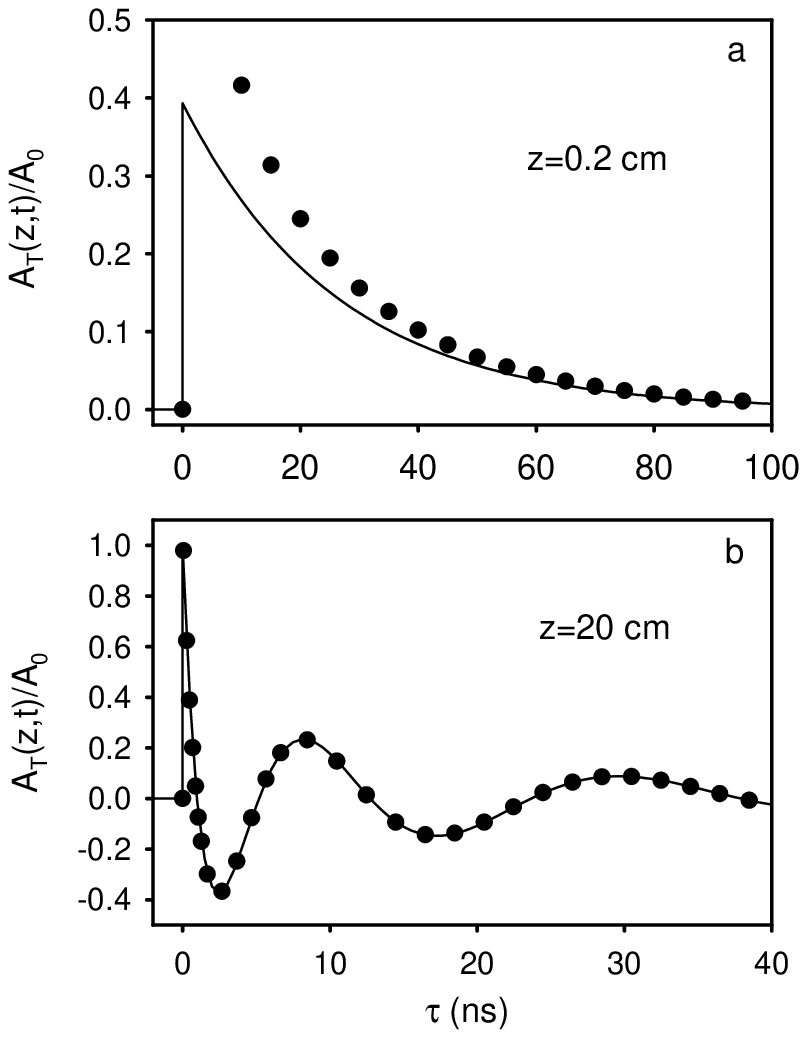}
\caption{\label{fig:comparison} The envelope for the total transient
field with the medium parameters of Ref. \cite{jeong} and
a) $z=0.2$ cm and b) $z=20$ cm.  The solid line shows the
predictions of the exact theory and the dots show the
predictions of the asymptotic (precursor) theory.}
\end{figure}

Our method can handle comparatively high number of oscillator densities, 
for which the condition $\alpha_0\ll\w_0/c$ and, hence, the \textit{a priori}
assumption of the SVAA (slowly varying amplitude approximation) 
are no longer  valid. 
Clearly, local field effects play a significant role in the 
expression of the main field. On the other hand, local field effects are 
still negligible in the expressions for the precursors. Indeed, the separation 
between the saddle-points and the resonant frequency is sufficiently high  
and the value of the refractive index at the saddle-points 
remains essentially unaffected. One verifies \textit{a 
posteriori}, that the SVAA holds in the derived formulae for the precursor fields, 
both in time and in space. Indeed, the separation of the scales of the carrier and beat frequencies is guaranteed by the relation $\partial q/\partial\tau\ll \w_0$ 
(equivalent to constraint \eqref{subd1}) and slow amplitude variation in time 
is guaranteed by $\delta\ll \w_0$. The small variation of amplitude in space
is guaranteed by  relation 
$(1/(2q))\partial q/\partial z\ll 2\partial q/\partial z$. 
(equivalent to $1\ll q$). The left side of this relation is exponential attenuation, obtained from bringing the rational attenuation to the exponent as a logarithm and taking the spatial derivative. Our results are consistent, 
in the sense that the sum of precursors 
agrees with the transient field obtained from the exact evaluation 
of the approximate field in Eq. \eqref{Eapprox} (see Fig. \ref{fig:comparison}). 
Our exact expression,  in terms of a series of Bessel 
functions, is  valid in  the higher density regime as well and agrees 
with Crisp's formula when restricted to low densities (see discussion below). 

We now give an example of the predictions of the precursor theory
and compare the results to the exact calculation of the
field for the dilute narrowband dielectric. In particular, we use the parameters of the experiment of Jeong \etal \ \cite{jeong} with
 $\w_0=2.5\times10^{15}$ $s^{-1}$, $\w_p=3\times10^9$ $s^{-1}$, $\delta=3\times10^7$ $s^{-1}$, $\w_c=\w_0$. We first consider the
case when the medium length is 0.2 cm, the value used in the experiment.  For these parameters, relation \eqref{largeq} guaranteeing a large value 
of $q$, becomes an equality ($q$ takes the value $q=1$) 
  at $\tau\approx 67$ ns, indicating that it takes on the order of hundreds of nanoseconds to satisfy the requirement. 
  Figure \ref{fig:comparison}a
compares the transient field envelopes for the two theories. While there is some discrepancy at shorter times, the error
is less than 25\% for times greater than 30 ns, and very small at $67$ ns, indicating that condition \eqref{largeq} is rather conservative in this case.
Thus, the primary contribution
to the transient field is from the saddle points and hence it is
reasonable to conclude that the experiments
of Jeong \textit{et al.} observed optical precursors. We note that they found that the exact theory agrees very well with the
experimental observations.

Figure \ref{fig:comparison}b compares the
two theories for a medium length 100 times longer than that used in the experiment ($z$=20 cm), where
it is seen that the agreement between them is excellent.
The increasing-period oscillations in the transient
field is clearly evident and consistent with our discussion above.
In this case, 
$q=1$ at $\tau\approx 6.7$ ns; the good agreement at as low a value as $q=1$ 
again points to the conservatism of the condition
 \eqref{largeq}. 

\section{Discussion}
\label{Sec:Discussion}

Crisp found that the step-modulated
input field evolves toward a so-called $0\pi$ pulse whose
envelope oscillates. (Note that a step-modulated incident
field inherently violates the SVAA, yet our solutions are
valid for this situation.)  He showed that the pulse area of
the total field  approaches zero, which is known as
a $0\pi$ pulse in the quantum optics community.
Such pulses have been studied experimentally by a number
of groups, beginning with the observation of
Rothenberg \textit{et al.}  \cite{rothenberg}, later work
demonstrating $0\pi$-pulse `stacking'  \cite{segard},
and more recent work \cite{matusovsky,sweetser,dudovich}.
Crisp posed the question of whether
 these weak-field coherent optical transients
(the $0\pi$ pulse) are a manifestation
of optical precursors and answered it
in the negative without mathematical proof. 
Certainly, Crisp explained these oscillations as an
interference between the parts of the pulse spectrum above
and below the atomic resonance frequency, where the
central part of the spectrum is eaten away as the field
propagates through the material.  He did not
associate these frequency components with the frequencies
of the Sommerfeld and Brillouin precursors, reasoning that 
precursors are an ultrafast effect, which would violate the
assumption of a slowly-varying amplitude, and thus
must be precluded from the SVAA formalism. Our precise
mathematical analysis proves conclusively that Crisp's
conclusion is incorrect. It follows clearly from our analysis 
that the oscillations
in the envelope of the $0\pi$ pulse is the result of the
interference of the Sommerfeld and Brillouin precursors.

Avenel \textit{et al.} \cite{avenel}, on the other
hand, first suggested that the coherent optical transients predicted by
Crisp and observed by Rothenberg \textit{et al.}~\cite{rothenberg}
are a manifestation of optical precursors and that the time scale for the precursors
can be very long (of the order of nanoseconds) for a
material with a narrow resonance.  However, they did
not provide a mathematical justification for their claim.  In
later work, Varoquaux \textit{et al.}~\cite{Varoquaux} attempted
to make their claim precise by solving Eq.\ (\ref{eq:E}) using
an asymptotic method.  They found a solution to the integral
only for frequencies well above the frequency of the
material resonance $\omega_0$
(see the discussion near the end of their Sec.\ IV.D.).
They predict that the envelope of the Sommerfeld precursor
contains oscillations similar in form to that predicted by
Crisp, which is the same as our exact solution (Eq.\ (\ref{exact})).
They were not able to identify a Brillouin precursor.  It
is not surprising that they failed to obtain an accurate
prediction concerning the precursors because the saddle
points are located close to $\omega_0$ (with respect to the 
scale of $\w_0$), yet
their approximate solution to the integral only accounted
for the contributions to the integral at much higher
frequencies. Our calculation identifies all the saddle-point 
contributions to the integral and places the Avenel \textit{et al.}
conjecture on a firm theoretical foundation.

In addition to the calculation of the precursors, we calculated
the total propagated field through the exact evaluation of the
simplified Fourier integral \eqref{Eapprox}. The calculation gives the total propagated field as a series of Bessel functions. This result was first obtained 
by Crisp \cite{crisp} under the additional assumption of the slowly 
varying amplitude approximation (SVAA). In the SVAA approach,
the wave equation is first simplified by assuming
a slowly-varying amplitude, then approximations
about the material are invoked, and a solution is
thus obtained. The use of a step-modulated nature of the 
initial field  in the context of the SVAA has raised questions. 
The exact agreement of Crisp's formula with our results (evaluated for a low density of oscillators), which makes no assumption of slowly varying amplitudes, demonstrates that, in spite of the initial 
discontinuity, the slowly-varying
assumption is superficial for a weak-field
and a narrow-resonance dilute medium.  In fact, it can be shown
that the solution to the SVAA equations for these
conditions also is a solution to the full wave equation. Our derivation 
shows the correct way to extend Crisp's formula to the high-density regime. 

Finally, we remark that our work also has implications
for very weak `quantum' fields.  Very recently,
Du \textit{et al.}~\cite{du,du2} have shown that precursors
can be observed on long time scales in correlated
bi-photon states.  A follow up study considering
the propagation of a classical field through a similar
medium has also been presented \cite{jeongdu}.

\begin{acknowledgements}
 
We thank Heejeong Jeong for useful discussions of this work and SV gratefully
acknowledges the support of NSF grants DMS-0207262 and DMS-0707488.

\end{acknowledgements}

\section{Appendix: Higher-Order Corrections }
\label{Sec:Corrections}

When the relation \be \psi_{rem}\ll q(\eta+\eta^{-1}) \ee
does not apply comfortably, we cannot neglect $\psi_{rem}$ altogether in determining the saddle-points. Instead, we correct our calculation of each precursor by retaining 
the linear Taylor approximation of $\psi_{rem}$ about the corresponding saddle point.  At the Sommerfeld saddle-point $\eta=1$, for example, the linear Taylor approximation of $\psi_{rem}$ is given by  
\be  \psi_{rem}=\g_0+\g_1\eta_1,\ee
where $\eta=1+\eta_1$, and  $\g_0$ and $\g_1$ are the values of $\psi_{rem}$ and its first derivative at the Sommerfeld saddle-point, respectively. 
After letting $\eta^{-1}=(1+\eta_1)^{-1}\approx 1-\eta_1+\eta_1^2$, 
we obtain for the exponent in \eqref{Eapprox}
\be -q(\eta+\frac{1}{\eta})\approx -q\left(2+\eta_1^2+\frac{\g_0}{q}+\frac{\g_1}{q}\eta_1,
\right),\ee 
for the stationary point 
\be \eta_1=-\frac{\g_1}{2q}, \ee 
for the exponent maximum
\be  -2q-\gamma_0+\frac{\gamma_1^2}{4q},\ee
and the second derivative at the stationary point
\be  -2q. \ee
In order to insert the corrections to the precursor field \eqref{somprec1}, we
\begin{enumerate}
\item Adjust the amplitude by multiplying the field  by the factor 
\[ \mbox{e}^{-\g_0+\g_1^2/4q}, \]
which inserts the correction of the exponent maximum  
(the second derivative of the exponent at 
the stationary point whose square root enters the fromula for 
the saddle-point contribution remains unchnaged in corrected exponent).
\item Perform the replacement 
\[(1-i\sigma) \rightarrow \left(1 -\frac{\g_1}{2q}-i\sigma\right), \]
in the denominator.
\end{enumerate}
The calculation is similarly straightforward for the 
Brillouin precursor. The corrective procedure maybe iterated by 
using the updated saddle-point as the base point. 

\bibliography{LeFew}

\begin{thebibliography}{22}
\expandafter\ifx\csname natexlab\endcsname\relax\def\natexlab#1{#1}\fi
\expandafter\ifx\csname bibnamefont\endcsname\relax
  \def\bibnamefont#1{#1}\fi
\expandafter\ifx\csname bibfnamefont\endcsname\relax
  \def\bibfnamefont#1{#1}\fi
\expandafter\ifx\csname citenamefont\endcsname\relax
  \def\citenamefont#1{#1}\fi
\expandafter\ifx\csname url\endcsname\relax
  \def\url#1{\texttt{#1}}\fi
\expandafter\ifx\csname urlprefix\endcsname\relax\def\urlprefix{URL }\fi
\providecommand{\bibinfo}[2]{#2}
\providecommand{\eprint}[2][]{\url{#2}}

\bibitem[{\citenamefont{Brillouin}(1960)}]{BrillouinBook}
\bibinfo{author}{\bibfnamefont{L.}~\bibnamefont{Brillouin}},
  \emph{\bibinfo{title}{Wave Propagation and Group Velocity}}
  (\bibinfo{publisher}{Academic Press}, \bibinfo{address}{New York},
  \bibinfo{year}{1960}).

\bibitem[{\citenamefont{Oughstun and Sherman}(1994)}]{OS}
\bibinfo{author}{\bibfnamefont{K.~E.} \bibnamefont{Oughstun}} \bibnamefont{and}
  \bibinfo{author}{\bibfnamefont{G.~C.} \bibnamefont{Sherman}},
  \emph{\bibinfo{title}{Electromagnetic Pulse Propagation in Causal
  Dielectrics}} (\bibinfo{publisher}{Springer-Verlag},
  \bibinfo{address}{Berlin}, \bibinfo{year}{1994}).

\bibitem[{\citenamefont{Choi and {\"O}sterberg}(2004)}]{choi}
\bibinfo{author}{\bibfnamefont{S.-H.} \bibnamefont{Choi}} \bibnamefont{and}
  \bibinfo{author}{\bibfnamefont{U.}~\bibnamefont{{\"O}sterberg}},
  \bibinfo{journal}{Phys. Rev. Lett.} \textbf{\bibinfo{volume}{92}},
  \bibinfo{pages}{193903} (\bibinfo{year}{2004}).

\bibitem[{\citenamefont{Albanese et~al.}(1989)\citenamefont{Albanese, Penn, and
  Medina}}]{bio_precursor}
\bibinfo{author}{\bibfnamefont{R.}~\bibnamefont{Albanese}},
  \bibinfo{author}{\bibfnamefont{J.}~\bibnamefont{Penn}}, \bibnamefont{and}
  \bibinfo{author}{\bibfnamefont{R.}~\bibnamefont{Medina}},
  \bibinfo{journal}{J. Opt. Soc. Am. B} \textbf{\bibinfo{volume}{6}},
  \bibinfo{pages}{1441} (\bibinfo{year}{1989}).

\bibitem[{\citenamefont{Roberts}(2004)}]{roberts}
\bibinfo{author}{\bibfnamefont{T.}~\bibnamefont{Roberts}},
  \bibinfo{journal}{Phys. Rev. Lett.} \textbf{\bibinfo{volume}{93}},
  \bibinfo{pages}{269401} (\bibinfo{year}{2004}).

\bibitem[{\citenamefont{Alfano et~al.}(2005)\citenamefont{Alfano, Birman, Ni,
  Alrubaiee, and Das}}]{alfano}
\bibinfo{author}{\bibfnamefont{R.}~\bibnamefont{Alfano}},
  \bibinfo{author}{\bibfnamefont{J.}~\bibnamefont{Birman}},
  \bibinfo{author}{\bibfnamefont{X.}~\bibnamefont{Ni}},
  \bibinfo{author}{\bibfnamefont{M.}~\bibnamefont{Alrubaiee}},
  \bibnamefont{and} \bibinfo{author}{\bibfnamefont{B.}~\bibnamefont{Das}},
  \bibinfo{journal}{Phys. Rev. Lett.} \textbf{\bibinfo{volume}{94}},
  \bibinfo{pages}{239401} (\bibinfo{year}{2005}).

\bibitem[{\citenamefont{Gibson and {\"O}sterberg}(2005)}]{gibson}
\bibinfo{author}{\bibfnamefont{U.}~\bibnamefont{Gibson}} \bibnamefont{and}
  \bibinfo{author}{\bibfnamefont{U.}~\bibnamefont{{\"O}sterberg}},
  \bibinfo{journal}{Opt. Express} \textbf{\bibinfo{volume}{13}},
  \bibinfo{pages}{2105} (\bibinfo{year}{2005}).

\bibitem[{\citenamefont{Avenel et~al.}(1984)\citenamefont{Avenel, Varoquaux,
  and Williams}}]{avenel}
\bibinfo{author}{\bibfnamefont{O.}~\bibnamefont{Avenel}},
  \bibinfo{author}{\bibfnamefont{E.}~\bibnamefont{Varoquaux}},
  \bibnamefont{and} \bibinfo{author}{\bibfnamefont{G.~A.}
  \bibnamefont{Williams}}, \bibinfo{journal}{Phys. Rev. Lett.}
  \textbf{\bibinfo{volume}{53}}, \bibinfo{pages}{2058} (\bibinfo{year}{1984}).

\bibitem[{\citenamefont{Jeong et~al.}(2006)\citenamefont{Jeong, Dawes, and
  Gauthier}}]{jeong}
\bibinfo{author}{\bibfnamefont{H.}~\bibnamefont{Jeong}},
  \bibinfo{author}{\bibfnamefont{A.~M.~C.} \bibnamefont{Dawes}},
  \bibnamefont{and} \bibinfo{author}{\bibfnamefont{D.~J.}
  \bibnamefont{Gauthier}}, \bibinfo{journal}{Phys. Rev. Lett.}
  \textbf{\bibinfo{volume}{96}}, \bibinfo{pages}{143901}
  (\bibinfo{year}{2006}).

\bibitem[{\citenamefont{Okawachi et~al.}(2007)\citenamefont{Okawachi, Slepkov,
  Agha, Geraghty, and Gaeta}}]{okawachi}
\bibinfo{author}{\bibfnamefont{Y.}~\bibnamefont{Okawachi}},
  \bibinfo{author}{\bibfnamefont{A.~D.} \bibnamefont{Slepkov}},
  \bibinfo{author}{\bibfnamefont{I.~H.} \bibnamefont{Agha}},
  \bibinfo{author}{\bibfnamefont{D.~F.} \bibnamefont{Geraghty}},
  \bibnamefont{and} \bibinfo{author}{\bibfnamefont{A.~L.} \bibnamefont{Gaeta}},
  \bibinfo{journal}{J. Opt. Soc. Am. A} \textbf{\bibinfo{volume}{24}},
  \bibinfo{pages}{3343} (\bibinfo{year}{2007}).

\bibitem[{\citenamefont{Crisp}(1970)}]{crisp}
\bibinfo{author}{\bibfnamefont{M.~D.} \bibnamefont{Crisp}},
  \bibinfo{journal}{Phys. Rev. A} \textbf{\bibinfo{volume}{1}},
  \bibinfo{pages}{1604} (\bibinfo{year}{1970}).

\bibitem[{\citenamefont{Oughstun and
  Cartwright}(2003{\natexlab{a}})}]{Oughstun2003}
\bibinfo{author}{\bibfnamefont{K.}~\bibnamefont{Oughstun}} \bibnamefont{and}
  \bibinfo{author}{\bibfnamefont{N.}~\bibnamefont{Cartwright}},
  \bibinfo{journal}{Opt. Express} \textbf{\bibinfo{volume}{11}},
  \bibinfo{pages}{1541} (\bibinfo{year}{2003}{\natexlab{a}}).

\bibitem[{\citenamefont{Oughstun and
  Cartwright}(2003{\natexlab{b}})}]{Oughstun2003a}
\bibinfo{author}{\bibfnamefont{K.}~\bibnamefont{Oughstun}} \bibnamefont{and}
  \bibinfo{author}{\bibfnamefont{N.}~\bibnamefont{Cartwright}},
  \bibinfo{journal}{Opt. Express} \textbf{\bibinfo{volume}{11}},
  \bibinfo{pages}{2791} (\bibinfo{year}{2003}{\natexlab{b}}).

\bibitem[{\citenamefont{Rothenberg et~al.}(1984)\citenamefont{Rothenberg,
  Grischkowsky, and Balant}}]{rothenberg}
\bibinfo{author}{\bibfnamefont{J.}~\bibnamefont{Rothenberg}},
  \bibinfo{author}{\bibfnamefont{D.}~\bibnamefont{Grischkowsky}},
  \bibnamefont{and} \bibinfo{author}{\bibfnamefont{A.}~\bibnamefont{Balant}},
  \bibinfo{journal}{Phys. Rev. Lett.} \textbf{\bibinfo{volume}{53}},
  \bibinfo{pages}{552} (\bibinfo{year}{1984}).

\bibitem[{\citenamefont{S{\'e}gard et~al.}(1987)\citenamefont{S{\'e}gard,
  Zemmouri, and Macke}}]{segard}
\bibinfo{author}{\bibfnamefont{B.}~\bibnamefont{S{\'e}gard}},
  \bibinfo{author}{\bibfnamefont{J.}~\bibnamefont{Zemmouri}}, \bibnamefont{and}
  \bibinfo{author}{\bibfnamefont{B.}~\bibnamefont{Macke}},
  \bibinfo{journal}{Europhys. Lett.} \textbf{\bibinfo{volume}{4}},
  \bibinfo{pages}{47} (\bibinfo{year}{1987}).

\bibitem[{\citenamefont{Matusovsky et~al.}(1996)\citenamefont{Matusovsky,
  Vaynberg, and Rosenbluh}}]{matusovsky}
\bibinfo{author}{\bibfnamefont{M.}~\bibnamefont{Matusovsky}},
  \bibinfo{author}{\bibfnamefont{B.}~\bibnamefont{Vaynberg}}, \bibnamefont{and}
  \bibinfo{author}{\bibfnamefont{M.}~\bibnamefont{Rosenbluh}},
  \bibinfo{journal}{J. Opt. Soc. Am. B} \textbf{\bibinfo{volume}{13}},
  \bibinfo{pages}{1994} (\bibinfo{year}{1996}).

\bibitem[{\citenamefont{Sweetser and Walmsley}(1996)}]{sweetser}
\bibinfo{author}{\bibfnamefont{J.}~\bibnamefont{Sweetser}} \bibnamefont{and}
  \bibinfo{author}{\bibfnamefont{I.}~\bibnamefont{Walmsley}},
  \bibinfo{journal}{J. Opt. Soc. Am. B} \textbf{\bibinfo{volume}{13}},
  \bibinfo{pages}{601} (\bibinfo{year}{1996}).

\bibitem[{\citenamefont{Dudovich et~al.}(2002)\citenamefont{Dudovich, Oron, and
  Silberberg}}]{dudovich}
\bibinfo{author}{\bibfnamefont{N.}~\bibnamefont{Dudovich}},
  \bibinfo{author}{\bibfnamefont{D.}~\bibnamefont{Oron}}, \bibnamefont{and}
  \bibinfo{author}{\bibfnamefont{Y.}~\bibnamefont{Silberberg}},
  \bibinfo{journal}{Phys. Rev. Lett.} \textbf{\bibinfo{volume}{88}},
  \bibinfo{pages}{123004} (\bibinfo{year}{2002}).

\bibitem[{\citenamefont{Varoquaux et~al.}(1986)\citenamefont{Varoquaux,
  Williams, and Avenel}}]{Varoquaux}
\bibinfo{author}{\bibfnamefont{E.}~\bibnamefont{Varoquaux}},
  \bibinfo{author}{\bibfnamefont{G.~A.} \bibnamefont{Williams}},
  \bibnamefont{and} \bibinfo{author}{\bibfnamefont{O.}~\bibnamefont{Avenel}},
  \bibinfo{journal}{Phys. Rev. B} \textbf{\bibinfo{volume}{34}},
  \bibinfo{pages}{7617} (\bibinfo{year}{1986}).

\bibitem[{\citenamefont{Du et~al.}(2008{\natexlab{a}})\citenamefont{Du,
  Kolchin, Belthangady, Yin, and Harris}}]{du}
\bibinfo{author}{\bibfnamefont{S.}~\bibnamefont{Du}},
  \bibinfo{author}{\bibfnamefont{P.}~\bibnamefont{Kolchin}},
  \bibinfo{author}{\bibfnamefont{C.}~\bibnamefont{Belthangady}},
  \bibinfo{author}{\bibfnamefont{G.}~\bibnamefont{Yin}}, \bibnamefont{and}
  \bibinfo{author}{\bibfnamefont{S.~E.} \bibnamefont{Harris}},
  \bibinfo{journal}{Phys. Rev. Lett.} \textbf{\bibinfo{volume}{100}},
  \bibinfo{pages}{183603} (\bibinfo{year}{2008}{\natexlab{a}}).

\bibitem[{\citenamefont{Du et~al.}(2008{\natexlab{b}})\citenamefont{Du,
  Belthangady, Kolchin, Yin, and Harris}}]{du2}
\bibinfo{author}{\bibfnamefont{S.}~\bibnamefont{Du}},
  \bibinfo{author}{\bibfnamefont{C.}~\bibnamefont{Belthangady}},
  \bibinfo{author}{\bibfnamefont{P.}~\bibnamefont{Kolchin}},
  \bibinfo{author}{\bibfnamefont{G.~Y.} \bibnamefont{Yin}}, \bibnamefont{and}
  \bibinfo{author}{\bibfnamefont{S.~E.} \bibnamefont{Harris}},
  \bibinfo{journal}{Opt. Lett.} \textbf{\bibinfo{volume}{33}},
  \bibinfo{pages}{2149} (\bibinfo{year}{2008}{\natexlab{b}}).

\bibitem[{\citenamefont{Jeong and Du}(2009)}]{jeongdu}
\bibinfo{author}{\bibfnamefont{H.}~\bibnamefont{Jeong}} \bibnamefont{and}
  \bibinfo{author}{\bibfnamefont{S.}~\bibnamefont{Du}}, \bibinfo{journal}{Phys.
  Rev. A} \textbf{\bibinfo{volume}{79}}, \bibinfo{pages}{011802}
  (\bibinfo{year}{2009}).

\end{thebibliography}

\end{document}